\documentstyle[prl,aps,twocolumn,floats,epsf]{revtex}
\begin{document}
\draft
\def\N{\mathbb{N}}
\def\R{\mathbb{R}}
\def\E{\mathbb{E}}
\def\P{\mathbb{P}}
\def\Ord{\mathcal{O}}
\title{Universal Distributions for Growth Processes in 1+1
  Dimensions\\ and Random Matrices}

\author{Michael Pr\"ahofer and Herbert Spohn}

\address{Zentrum Mathematik and Physik Department, TU M\"unchen, 
  D-80290 M\"unchen, Germany.\\ Email: \emph{praehofer@ma.tum.de,
    spohn@ma.tum.de
    }}

\date{\today}


\maketitle
\begin{abstract}
We develop a scaling theory for KPZ growth in one
dimension by 
a detailed study of the polynuclear growth (PNG) model. In particular,
we identify three universal distributions for shape fluctuations and
their dependence on the macroscopic shape.
These distribution functions are computed using the partition function
of Gaussian random matrices in a cosine potential.
\end{abstract}

Growth processes lead to a rich variety of macroscopic patterns and
shapes \cite{meakin}. As has been recognized for some time, growth
may also give rise to intriguing statistical fluctuations
comparable to 
thermal fluctuations at a critical point. One of the most prominent
examples is the Kardar-Parisi-Zhang (KPZ) universality
class \cite{KPZ86}. In
essence one models a stable phase which grows into an unstable phase
through aggregation, as for example in Eden type models where
perimeter sites of a given cluster are filled up randomly. In real
materials, mere aggregation is often too simplistic an assumption
and one would have to take other dynamical modes, such as surface
diffusion, at the stable/unstable interface into account
\cite{krug}.  In our letter we remain within the KPZ class.

From the beginning there has been evidence that in one spatial
dimension KPZ growth processes are linked to exactly soluble models
of two-dimensional statistical mechanics. Kardar \cite{kardar}
mapped growth to the directed polymer problem. The replica trick
then yields the Bose gas with attractive $\delta$--interaction which
in one dimension can be solved through the Bethe ansatz \cite{LL}.
In \cite{gwaspohn}, considerably generalized in \cite{kim}, for a
particular discrete growth model the statistical weights for the
local slopes were mapped onto the six vertex model.  
To solve the six vertex model one
diagonalizes the transfer matrix, again, through the Bethe ansatz,
which also 
allows for a study of finite size scaling \cite{DLA}.
Unfortunately none of these methods go beyond what corresponds to
the free energy in the six vertex model and
the associated dynamical scaling exponent $\beta=1/3$.

In this letter we point out that within the KPZ universality class
the polynuclear growth (PNG) model plays a distinguished role: it
maps onto random permutations, the height being the length of the
longest increasing subsequence of such a permutation, and thereby
onto Gaussian random matrices \cite{rains,AD99}. 
We use these mappings to 
obtain an analytic expression for certain scaling
distributions, which then leads to an understanding of how the
self-similar height fluctuations depend on the initial conditions
and to a more refined scaling theory for KPZ growth.

PNG is a simplified model for layer by layer growth \cite{meakin}. 
One starts with a perfectly flat crystal in contact with its
super-saturated vapor. Once in a while a supercritical nucleus is formed,
which then spreads laterally by further attachment of
particles at its perimeter sites. Such islands coalesce if they are in
the same layer and further islands may be nucleated upon already existing
ones. The PNG model ignores the lateral lattice structure and assumes
that the 
islands are circular and spread at constant speed. The nucleation
rate and the lateral speed can be set to one by the appropriate choice
of space-time units.


We specialize to a one-dimensional surface, returning to 
higher dimensions at the end. The height, $h(x,t)$, at time $t$ above
the point $x$ on the substrate is counted in lattice spacings. 
The upward steps of $h$ move deterministically with velocity $-1$, the
downward steps with velocity $+1$, and they annihilate upon
touching. Through a nucleation event at $(x,t)$, randomly in
space-time, $h$ increases at $x$ by one unit thereby creating a new
up-down pair of steps.
To explain the mapping from PNG to 
permutations it is convenient to first use a droplet geometry, 
where a single island starts spreading from the 
origin and further nucleations take place only above this
ground layer.
The initially flat substrate and other initial conditions will be
handled along the lines of this blueprint. 

\begin{figure}[ht]
  \begin{center}
    \mbox{\epsfxsize=7cm\epsffile{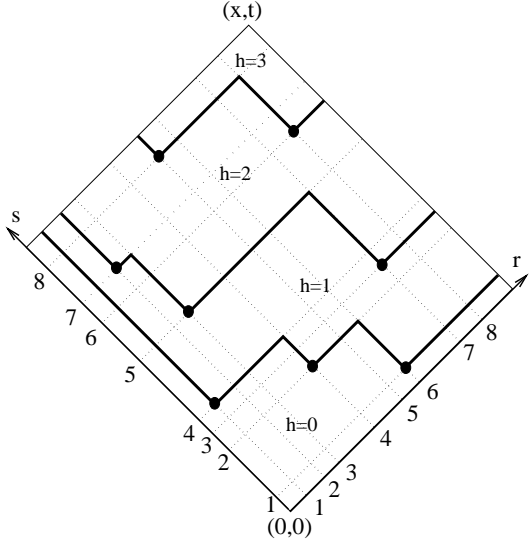}}
    \vspace{3mm}
    \caption{The height $h$ of a PNG droplet with nucleation
      events corresponding to the permutation $(4,7,5,2,8,1,3,6)$.}
    \label{fig:droplet}
  \end{center}
\vspace{-9mm}
\end{figure}

We want to compute the height $h(x,t)$ of the droplet. Clearly it is
determined by 
the set of nucleation events inside the rectangle
$R_{(x,t)}=\{(x',t'):|x'|\leq t'\mbox{ and }|x-x'|\leq t-t'\}$.  In
lightlike coordinates, $r=(t'+x')/\sqrt{2}$, $s=(t'-x')/\sqrt{2}$,
the rectangle $R_{(x,t)}$ equals $[0,R]\times[0,S]$ with
$R=(t+x)/\sqrt{2}$, $S=(t-x)/\sqrt{2}$.  We label the nucleation
events as $(r_n,s_n)$, $n=1,\ldots,N$, such that $0\leq
r_1<\cdots<r_N\leq R$. The corresponding order in the second
coordinate $s$, $0\leq s_{p(1)}<\cdots<s_{p(N)}\leq V$, defines then
a permutation $p$ of length $N$,
compare with Fig.~\ref{fig:droplet}.

There is a simple rule of how to determine the number of the layer in
which each nucleation event is located. 
Points in layer $1$ are obtained by scanning
the permutation $(p(1),\ldots,p(N))$ from left to right and marking
all those entries which are smaller than the so far smallest.  
After deleting the subsequence of the first layer,
the second layer is obtained by repeating this construction. One
marks those of the
remaining entries of the permutation which are in decreasing order.
At the end the permutation $p$ has been subdivided into decreasing
subsequences.
In the example of Fig. \ref{fig:droplet} we have the permutation
$(4,7,5,2,8,1,3,6)$. The first decreasing subsequence is $(4,2,1)$,
corresponding to the nucleation events in the first layer. The
remaining subsequences are $(7,5,3)$ and $(8,6)$.
The height $h(x,t)$ is the number of
these subsequences 
and therefore the length of the longest increasing subsequence of $p$
\cite{AD99}.

In a dual picture one draws a directed path from $(0,0)$ to
$(x,t)$, joining nucleation events by straight lines, with the
restriction that 
both coordinates $r$ and $s$ are increasing along the path.
Equivalently, the path must be in the forward light cone at each
nucleation event.  This is the celebrated directed polymer, cf. for
example \cite{meakin}, in the
context of the PNG model. If to each path we assign as
negative energy 
the number of nucleation centers traversed, then $h(x,t)$ equals the
ground state energy of the directed polymer. The PNG model is thus
in the strong coupling regime.

The nucleation events have density one and are independently and
uniformly distributed in 
the rectangle $R_{(x,t)}$ with area $\lambda=(t^2-x^2)/2$. This 
induces a Poisson distribution for the length, $N$, of the
permutation, $\mbox{Prob}\{N=n\}=e^{-\lambda}\lambda^n/n!$, and for a
given length $n$ each permutation has the same probability, namely
$1/n!$. Thus the problem of computing the distribution of the height
$h(x,t)$ is converted into determining the statistics of the length, $l$,
of a longest increasing subsequence of a random permutation. Since to
leading order $h(x,t)\propto t$, $l$ must be of order
$\sqrt{\lambda}$ and the relative fluctuations should be of order
$\lambda^{1/6}$, if we accept $\beta=1/3$ 
for KPZ growth in $1+1$ dimension.

The same construction can be carried out for an initially flat
substrate.  By translation invariance, it suffices to study
$H(t)=h(0,t)$.  The rectangle $R_{(0,t)}$ is now replaced by the
triangle $T_t=\{(x',t'):|x'|\leq t-t', \,\,t'\geq0\}$. To relate to
the directed polymer we add the mirror image relative to $t=0$,
including the nucleation events, to obtain the square
$R_{t}=\{(x',t'):|x'|\leq t-|t'|\}$. Then $2H(t)$ equals again the
ground state energy of the directed polymer from $(0,-t)$ to $(0,t)$.
However the statistics of nucleation centers inside $R_t$ is
constrained to satisfy the reflection symmetry relative to $t=0$.

For a random permutation with Poisson distributed length $N$, $\langle
N\rangle=\lambda$, the length $l$ of the longest
increasing subsequence satisfies the amazing identity
\begin{equation}
  \label{eq:finite}
  \mbox{Prob}\{l\leq m\}=e^{-\lambda}\int_{m\times m}dU
  \exp\left({\sqrt{\lambda}\,\mbox{Tr}(U+U^{-1})}\right),
\end{equation}
where the integration is uniformly over all $m\times m$ unitary
matrices. 
A proof can be found for example in \cite{AD99,BDJ98}.
The partition function in (\ref{eq:finite}) appeared before
in the context of quantum gravity and has a third order phase
transition at $m\simeq2\sqrt{\lambda}$ with finite size scaling
governed by the Painlev\'e II equation \cite{grosswitten,PS90}.
Baik \emph{et al}
\cite{BDJ98} prove that $l\simeq 2\sqrt{\lambda}+\lambda^{1/6}\chi_2$ for
large $\lambda$, where $\chi_2$ is a random variable 
distributed according to the GUE Tracy-Widom distribution, i.~e.~the
distribution of the largest eigenvalue of a complex hermitian random
matrix \cite{TW94}. One has $\mbox{Prob}\{\chi_2\leq
x\}=F_2(x)=e^{-g(x)}$, where
$g''(x)=\mbox{u}(x)^2$, $g(x)\to0$ as $x\to\infty$, and $\mbox{u}(x)$
is the global positive solution of the
Painlev\'e II equation $\mbox{u}''=2\mbox{u}^3+x\mbox{u}$. Its asymptotics
are $\mbox{u}(x)\sim\sqrt{-x/2}$ for $x\to-\infty$ and
$\mbox{u}(x)\sim\mbox{Ai}(x)$ for $x\to\infty$, $\mbox{Ai}(x)$ the
Airy function. 

To translate to the PNG model we introduce the growth velocity $v(u)$,
depending on the macroscopic slope $u=\partial h/\partial x$, and the
static roughness $A(u)$ \cite{PS99}, which for PNG are
$v(u)=\sqrt{2+u^2}$, $A(u)=\sqrt{2+u^2}$ in our units. Then
\begin{equation}
  \label{eq:dropletscaling}
  h(v'(u)t,t)\simeq \bbox{(}v(u)-uv'(u)\bbox{)}t
  +({\textstyle\frac12}v''(u)A(u)^2\,t)^{1/3}\chi_2
\end{equation}
in the limit of large $t$.
We emphasize that all nonuniversal factors are given through the model
dependent quantities 
$v(u)$, $A(u)$
and remark that
(\ref{eq:dropletscaling}) is also confirmed by the rigorous result
of Johansson \cite{J99} for a discrete growth model equivalent to the
totally asymmetric simple exclusion process.

\begin{figure}[htb]
  \begin{center}
    \mbox{\epsfxsize=8cm\epsffile{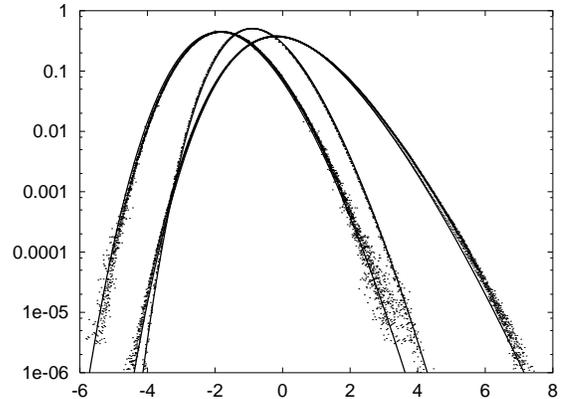}}
    \caption{From left to right: the probability densities of the
      universal distributions $\chi_2$, $\chi_1$, and $\chi_0$ for
      curved, flat, and stationary self-similar growth, respectively.}
    \label{fig:distributions}
  \end{center}
\vspace{-7mm}
\end{figure}

For the flat substrate one might expect to have the same fluctuation
law as for the droplet, since in both cases the mean curvature
vanishes on a microscopic scale. A result of Baik and Rains
\cite{BR99} tells 
us however that the fluctuations are GOE \cite{TW94}. 
More precisely, there is a 
similar formula for $\mbox{Prob}\{l\leq m\}$ as (\ref{eq:finite}) in
the case that the random permutation $p$ is reflection symmetric relative
to the anti-diagonal, $p\bbox{(}N+1-p(k)\bbox{)}=N+1-k$. The asymptotic
analysis of \cite{BR99} results
in $l\simeq4\sqrt{\lambda}+(2\lambda)^{1/6}\chi_1$, where $\chi_1$ is
distributed as the largest eigenvalue of a real symmetric random
matrix. Translated to the surface this means
\begin{equation}
  \label{eq:flatzeroscaling}
  H(t)=\sqrt{2}t+(t/\sqrt{2})^{1/3}\chi_1.
\end{equation}
One has $\mbox{Prob}\{2^{2/3}\chi_1\leq
x\}=F_1(2^{-2/3}x)=e^{-(f(x)+g(x))/2}$, 
$g(x)$ as above and
$f'(x)=-\mbox{u}(x)$, $f(x)\to0$ for $x\to\infty$.
The distributions
of $\chi_2$ and $\chi_1$ are plotted in
Fig.~\ref{fig:distributions}. Superimposed are Monte Carlo data for the
PNG model, which  
differ distinguishably from the analytical curves only at the tails
where statistics becomes bad.
We conclude
that the droplet and the flat substrate have the same scaling form but
distinct universal distributions.

The flat substrate, although used in many simulations, and the droplet
are rather special as initial conditions. From a statistical mechanics
point of view stationary growth would be regarded as singled out,
which for PNG corresponds to 
initial conditions where the up and down steps are random with
densities $\sqrt{2}$ each. Physically,
another natural initial condition is to have a staircase
configuration representing a tilted surface. In addition
we could have sources, for example additional nucleation events at the
origin. The
mapping to the directed polymer works as before. 
Our crucial observation is that such other initial conditions
translate in essence to 
defect lines and/or boundary
potentials for the directed polymer.

To illustrate, we discuss only one special geometry.
As for the droplet we consider
random nucleations of density one in the square $R_{(0,t)}$. In
addition there are random nucleations at the two lower edges $\{s=0\}$
and $\{r=0\}$ with constant line densities $\rho_+$, resp. $\rho_-$.
Thus the path of minimal energy, with starting  point at $(0,0)$,
sticks for a while at one of the two edges and then enters the bulk to
reach 
$(0,t)$ eventually. If $\rho_+<1$, $\rho_-<1$, it does not pay to stay
at the edges, and from the bulk we have GUE energy fluctuations
according to (\ref{eq:dropletscaling}). On the other hand if
$\rho=\max\{\rho_+,\rho_-\}>1$, the optimal path stays for a length
$t(1-1/\rho^2)/\sqrt{2}$ at the edge with the higher density. Since
the edge events are random, the $t^{1/3}$ bulk fluctuations are
dominated by the Gaussian $\sqrt{t}$ edge fluctuations. Parenthetically
we remark that for regularly spaced edge points one should recover
GOE. 

The length distribution along the critical lines $\rho_+=1$, $\rho_-<1$,
resp. $\rho_+<1$, $\rho_-=1$, was identified by Baik and Rains \cite{B},
in a generalization of the techniques in \cite{BR99}. They obtain
GOE$^2$ fluctuations, i.e. the distribution of the maximum 
of two independent GOE random variables. The path of
minimal energy stays for a 
length of order $t^{1/3}$ at the density one edge.
At the critical point $\rho_+=\rho_-=1$, the polymer has a choice
between the left and right edge. By a limiting procedure one
obtains \cite{B} the universal distribution for the energy fluctuations,
 $F_0(x)=\mbox{Prob}\{\chi_0<x\}$, with
\begin{equation}
  \label{eq:stationarydistr}
  F_0(x)=[1-(x+2f''+2g'')g']e^{-(g+2f)}.
\end{equation}
An interpretation in terms of the eigenvalue distribution of random
matrices has yet to be found. 
In Fig.~\ref{fig:distributions} we plot the distribution of
$\chi_0$. Superimposed are 
simulation data for the PNG model, taken before the analytic result
had been obtained. 
The first four moments of $\chi_j$, $j=0,1,2$, are listed in Table
\ref{tab:1}.
Of interest are also the asymptotics of the probability densities
$F_j'(x)$. From Painlev\'e II we obtain $-\log
F_j'(x)=c_j|x|^3/12$ for $x\to-\infty$ and $-\log
F_j'(x)=d_jx^{3/2}/3$ for $x\to\infty$ up to logarithmic
corrections with prefactors $c_j=1,2,1$ and $d_j=2,4,4$ for
$j=0,1,2$, respectively. 

\begin{table}[b]
  \caption{Mean, variance, skewness, and kurtosis for the
    distributions of $\chi_2$, $\chi_1$, and $\chi_0$ as determined by
    numerically solving Painlev\'e II [19]. 
    $\langle\chi^n\rangle_c$ denotes the $n$'th cumulant.}
  \label{tab:1}
  \begin{tabular}{cddd}
     & curved ($\chi_2$)& flat ($\chi_1$)&stationary ($\chi_0$)\\
    \tableline
    $\langle\chi\rangle$&$-$1.77109&$-$0.76007&$0$\\
    $\langle\chi^2\rangle_c$&0.81320&0.63805&1.15039\\
    $\langle\chi^3\rangle_c/\langle\chi^2\rangle_c^{3/2}$
     &0.2241&0.2935&0.35941\\ 
    $\langle\chi^4\rangle_c/\langle\chi^2\rangle_c^{2}$
    &0.09345&0.1652&0.28916\\
  \end{tabular}
\end{table}

\begin{figure}[htbp]
  \begin{center}
    \mbox{\epsfxsize=7cm\epsffile{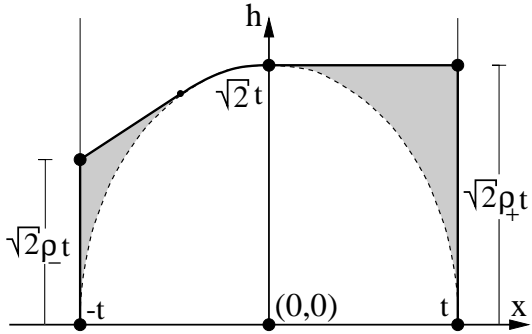}}
    \vspace{5mm}
    \caption{Droplet with boundary sources of intensity $\rho_-<1$
    and $\rho_+=1$ (critical). The dashed line corresponds to the
    free droplet. The shaded region is the extra mass
    due to sources.}
    \label{fig:stationary}
  \end{center}
\vspace{-7mm}
\end{figure}
We have to translate back to surface growth. For stationary growth
with zero slope, in the space-time picture, the height lines cross the
forward light cone with the densities $\rho_+=1=\rho_-$ and
the intersection points are
Poisson distributed \cite{AD95}. Thus for the
directed polymer with edge densities the critical point is precisely
stationary growth with zero slope. 
If $\rho_+\rho_-=1$, $\rho_+\neq1$, we have also stationary growth but
now with slope 
$u=(\rho_--\rho_+)/\sqrt{2}$. As argued already the fluctuations along
the line $x=0$ are then Gaussian $\sqrt{t}$. For the $t^{1/3}$
fluctuations one has to record height differences along the line
$\{x=v'(u)t\}$, as can be seen from a similarity transformation. In
Fig.~\ref{fig:stationary} we illustrate the macroscopic shape for general
boundary sources. If $\rho_+=0=\rho_-$, we have the droplet discussed
before. Nonzero boundary sources enforce flat segments tangential to the
droplet shape. The profile at $x=0$ is curved for  $\rho_+<1$,
$\rho_-<1$, flat otherwise, the marginal case corresponding to the
critical lines.

Our detailed study of the PNG model suggests the following scaling
theory for all growth 
models in the KPZ universality class. First of all we require a
self-similar macroscopic shape. Locally this leaves only two possibilities,
either a flat piece or a curved piece with a shape determined
through the slope dependent growth velocity \cite{KS92}. We draw a ray
from the 
center of symmetry. If the surface at the point of intersection with
the ray has non-zero 
curvature, then the height fluctuations in this direction are GUE with
scaling form (\ref{eq:dropletscaling}). If the curvature is zero, we
have to know the roughness of the initial
conditions, i.~e.~$|h(x,0)-h(0,0)|\propto|x|^\alpha$ with roughness
exponent $\alpha$, and/or the corresponding roughness for boundary
sources. If $\alpha=0$ the height fluctuations are GOE and the general
scaling form is as in (\ref{eq:dropletscaling}) with $\chi_2$ replaced
by $\chi_1$.
If $\alpha=1/2$ the height fluctuations are Gaussian with variance
proportional to $t$, except along the line $\{x=v'(u)t\}$, where they
again have the scaling form (\ref{eq:dropletscaling}) with the random
variable $\chi_2$ replaced by $\chi_0$, as defined in
(\ref{eq:stationarydistr}). The
intermediate cases $0<\alpha<\frac12$ have not been studied
systematically. Also the fluctuations at the endpoints of flat pieces
have still to be classified. There are two exceptions. One is the case
of Fig.~\ref{fig:stationary}, which has GOE$^2$ and the second one is
the half-droplet with an
external source at $x=0$. Translating \cite{BR99} to PNG one finds
GOE, GSE, and Gaussian depending on the strength of the source.

Our constructions carry over immediately to higher dimensions, as can
be seen most directly in the polymer picture. The square is replaced
by a $(d+1)$--dimensional (hyper)cube with uniformly distributed
nucleation centers, the polymer running from the lower to the upper
tip. For the PNG model this corresponds to droplet growth with islands
having the shape of a regular simplex (a triangle in $2+1$, a
tetrahedron in $3+1$, and so on). One axis of the cube defines the
order $1,\ldots,N$, while the remaining $d$ axes define the permutations
$p_i(1),\ldots,p_i(N)$, $i=1,\ldots,d$. Increasing means now
increasing in all coordinates, i.e. $j<j'$ and $p_i(j)<p_i(j')$ for
all $1\leq i\leq d$.  The length of the longest increasing subsequence
equals, again, the height of the droplet. At present we study
numerically the statistics of this length with the goal to have
information on scaling more precise than the one of previous investigations
\cite{tang}.

In conclusion, we have obtained distinct scaling functions for the PNG
model, which depend on the choice of initial conditions. By
universality we argued that from the knowledge of the self-similar
curvature one can infer the type of height fluctuations. It would be
of interest 
to study also joint probability distributions of the height at
distinct space-time points. Perhaps such a program could identify the
universal field theory hiding behind KPZ growth in one dimension.

We are grateful to J.~Baik and E.~M.~Rains for making their results \cite{B}
available to us prior to publication and to C. Tracy for help on
Painlev\'e II. 

\end{document}